\documentclass[preprint,showpacs,preprintnumbers,amsmath,amssymb]{revtex4}

\usepackage{graphicx}
\usepackage{dcolumn}
\usepackage{bm}
\usepackage{subfigure}
\usepackage{theorem}
\newcommand{\mib}[1]{\mbox{\boldmath $#1$}}

\def\R{{\bf R}}
\def\C{{\bf C}}
\def\N{{\bf N}}
\def\W{{\bf W}}

\def\K{{\bf K}}

\def\LE{{\cal L}}
\def\im{{\rm i}}

\def\n{{\bf n}}
\def\vphi{\mib{\varphi}}
\def\vrho{\mib{\rho}}
\def\vtheta{\mib{\theta}}
\def\vpi{\mib{\pi}}

\def\z{{\bf z}}
\def\w{{\bf w}}

\def\Re{{\rm Re}\,}
\def\Im{{\rm Im}\,}

\begin{document}

\preprint{}

\title{Determinantal Correlations of Brownian Paths in the Plane \\
with Nonintersection Condition on their Loop-Erased Parts}

\author{Makiko Sato}
\author{Makoto Katori}
\email{katori@phys.chuo-u.ac.jp}
\affiliation{%
Department of Physics,
Faculty of Science and Engineering,
Chuo University, 
Kasuga, Bunkyo-ku, Tokyo 112-8551, Japan 
}%

\date{9 March 2011}

\begin{abstract}
As an image of the many-to-one map of loop-erasing operation $\LE$
of random walks, a self-avoiding walk (SAW) is obtained.
The loop-erased random walk (LERW) model is the statistical
ensemble of SAWs such that the weight of each SAW $\zeta$
is given by the total weight of all random walks $\pi$ 
which are inverse images of $\zeta$,
$\{\pi: \LE(\pi)=\zeta \}$.
We regard the Brownian paths as the continuum limits
of random walks and consider the statistical ensemble of
loop-erased Brownian paths (LEBPs) as the continuum
limits of the LERW model.
Following the theory of Fomin on nonintersecting
LERWs, we introduce a nonintersecting system of
$N$-tuples of LEBPs in a domain $D$
in the complex plane, where the total weight
of nonintersecting LEBPs is given by Fomin's determinant
of an $N \times N$ matrix whose entries
are boundary Poisson kernels in $D$.
We set a sequence of chambers
in a planar domain and observe 
the first passage points at which $N$ Brownian paths
$(\gamma_1, \dots, \gamma_N)$ first enter each chamber, 
under the condition that the loop-erased parts
$(\LE(\gamma_1), \dots, \LE(\gamma_N))$ 
make a system of nonintersecting LEBPs in the domain
in the sense of Fomin.
We prove that the correlation functions of first passage points
of the Brownian paths of the present system 
are generally given by determinants
specified by a continuous function called the correlation kernel.
The correlation kernel is of Eynard-Mehta type,
which has appeared in two-matrix models
and time-dependent matrix models 
studied in random matrix theory.
Conformal covariance of correlation functions is 
demonstrated.

\end{abstract}

\pacs{05.40.-a,02.50.-r,02.30.Fn}

\maketitle

\section{Introduction}

The vicious walker model introduced by Fisher \cite{Fis84}
is a one-dimensional system of simple symmetric random walks
conditioned so that any pairs of trajectories of walkers
in the 1+1 dimensional spatio-temporal lattice
are nonintersecting.
The transition probability of $N$ vicious walkers
can be described by using a determinant of an $N \times N$
matrix, whose entries are transition probabilities
of a single random walker with different initial and
final positions. This determinantal expression
for nonintersecting paths is called
the Karlin-McGregor formula in probability theory
\cite{KM59} and the  Lindstr\"om-Gessel-Viennot
formula in enumerative combinatorics \cite{Lin73,GV85}.
Preserving the determinantal expression for transition
probability densities, a continuum limit (the diffusion scaling limit)
of vicious walkers can be taken and we have
the system of one-dimensional Brownian motions
conditioned never to collide with each other
\cite{KT02,KT03,CK03}. The important fact is that
the obtained interacting particle systems defined
in the continuous spatio-temporal plane,
which can be called the noncolliding Brownian motion \cite{KT07},
is identified with Dyson's Brownian motion model with
$\beta=2$ \cite{Dys62},
which was originally introduced as a stochastic
process of eigenvalues of an Hermitian-matrix valued
Brownian motion in the random matrix theory
\cite{Meh04,Forr10}. 
The notion of correspondence between 
nonequilibrium particle systems
and random matrix theories is very useful \cite{KT04}
and spatio-temporal correlation functions
of noncolliding diffusion processes have been
determined explicitly not only for the systems
with finite numbers of particles
but also for the systems with infinite numbers of
particles \cite{NKT03,KT07,KT09,KT10a,KT10b}.

In the present paper, we study a system of continuous paths
not in the 1+1 dimensional spatio-temporal plane
but in the two dimensional plane 
({\it i.e.} the complex plane $\C=\{z=x+\im y\}$
with $\im =\sqrt{-1}$),
which will be called the 
nonintersecting system of loop-erased Brownian paths (LEBPs)
\cite{KL05}.
A version of nonintersection condition is imposed
between the paths
(see Eq.(\ref{eqn:LEBP1}) below),
and then
the total weight of LEBPs is given by
Fomin's determinant \cite{For01,LL10}
instead of the determinant of Karlin-McGregor
(and of Lindstr\"om-Gessel-Viennot). 
There the entries of matrix whose determinant is
considered are the normal derivatives 
at boundary points of domain of
the Green's functions of the two-dimensional
Poisson equation 
(the Poisson kernels and the boundary Poisson kernels)
instead of the transition probability densities
\cite{KL05}.

In Section II, we define the LERW model and 
briefly review Fomin's theory
of nonintersecting LERWs.

As the continuum limit of LERW model, the statistical
ensemble of LEBPs is introduced in Section III.A.
There the Green's function, the Poisson kernel,
and the boundary Poisson kernel are defined for
the Brownian motion in a domain in the 
two-dimensional plane or the complex plane $\C$. 
Then for an $L \times \pi$ 
rectangular domain $R_{L}$ in $\C$, Fomin's
determinant of the boundary Poisson kernels
and of the Poisson kernels for $N$-tuples of
Brownian paths are studied, and nonintersecting
system of LEBPs are constructed in 
Section III.B (see Fig.\ref{fig:Fig3}).
In Section III.C, we set two rectangular domains on $\C$
adjacent to each other at a vertical line $\Re z=x$,
in which $N$ Brownian paths are running from the left
rectangular domain to the right one through the
line $\Re z=x$ (see Fig.\ref{fig:Fig4}). 
We impose the condition that the loop-erased parts
of Brownian paths are nonintersecting in the sense
of Fomin as expressed by Eq.(\ref{eqn:LEBP1}).
Under this condition, the probability density function of the
first passage points on the line $\Re z=x$, at which
the $N$ Brownian paths enter the right rectangular domain from 
the left domain, are given as 
Eqs.(\ref{eqn:pL1}) for $L < \infty$ 
and (\ref{eqn:pN1}) for $L \to \infty$, respectively.
In Section III.D, we consider a sequence of $M+1$
rectangular domains on $\C$, $M \in \N \equiv \{1,2, \dots\}$,
where the $m$-th domain and the $(m+1)$-th domain
are adjacent at the line $\Re z=x_m, 1 \leq m \leq M$.
$N$-tuples of Brownian paths are running from the left to the right
(see Fig.\ref{fig:Fig5}) under the condition that
their loop-erased parts make a nonintersecting system of LEBPs.
The probability density function of joint distributions
of first passage points at $M$ lines
$\Re z=x_m, 1 \leq m \leq M$, of the Brownian paths
are determined as Eqs.(\ref{eqn:jointpL}) for $L < \infty$
and (\ref{eqn:jointp}) for $L \to \infty$, respectively.

In Section IV, a special initial condition is
assumed when the $N$-tuples of Brownian paths 
start from the left boundary of the leftmost domain.
In this special case, we can explicitly obtain 
all multiple correlation functions of first passage points
on the lines $\Re z=x_m, 1 \leq m \leq M$
for any $M \in \N$,
in which they are given by determinants (Theorem 1).
The correlation kernel, which completely specifies
the determinants, are given by
Eqs.(\ref{eqn:K1}) and (\ref{eqn:K2}).
The statistical ensemble of points
whose correlation functions are generally expressed by 
determinants with a correlation kernel is
called a {\it determinantal point process}
or a {\it Fermion point process} 
in probability theory \cite{Sos00,ST03,HKPV09}.
It should be noted that the present correlation
kernel is asymmetric 
$\K^{\pi/2}_N(x, \theta; x', \theta')
\not= \K^{\pi/2}_N(x', \theta'; x, \theta)$
for $x \not= x'$ as shown by Eqs.(\ref{eqn:K1})
and (\ref{eqn:K2}). 
This asymmetric correlation kernel is 
of {\it Eynard-Mehta type} \cite{EM98},
which has been studied in two-matrix models
and time-dependent matrix models in 
random matrix theory \cite{Meh04,Forr10,KT02}.

Since the Brownian motions and their loop-erased parts
on $\C$ are {\it conformally invariant} \cite{KL05},
our correlation kernel is {\it conformally covariant}.
In Section V, the determinantal correlation functions
in the half-infinite-strip domain
$R \equiv \lim_{L \to \infty} R_L
=\{z \in \C: \Re z >0, 0 < \Im z < \pi\}$
given by Theorem 1 
is mapped to the domain
$\Omega=\{z = r e^{\im \theta} \in \C :
r>1, 0 < \theta < \pi\}$
(Corollary 2).
There numerical plots of the density function
and the two-point corrrelation functions
in the domain $\Omega$ are shown by figures.

Concluding remarks are given in
Section VI. 
Appendix A is prepared to derive the formulas
of the Poisson kernel and the boundary Poisson kernel
used in the text.

\section{Loop-Erased Random Walks and Fomin's Determinant}

We consider an undirected planar lattice
consisting of a set of vertices (sites)
$V=\{v_j\}$ and a set of edges (bonds)
$E=\{e_j\}$. 
Together with a set of the weight functions
of the edges $W=\{w(e)\}_{e \in E}$,
a network $\Gamma=(V,E,W)$ is defined.

For $a, b \in V$, 
let $\pi$ be a {\it walk} given by
\begin{equation}
\pi \, : \,
a=v_0 \,
{\stackrel{e_1}{\rightarrow}} \, 
v_1 \, 
{\stackrel{e_2}{\rightarrow}} \, 
v_2 \, 
{\stackrel{e_3}{\rightarrow}} \, 
\cdots
{\stackrel{e_m}{\rightarrow}} \, 
v_m =b \,
\label{eqn:aj}
\end{equation}
where the length of walk is 
$|\pi|=m \in \N$ and, 
for each $0 \leq j \leq m-1$, 
$v_j$ and $v_{j+1}$ are nearest-neighboring vertices
in $V$ and $e_j \in E$ is the edge connecting
these two vertices. 
We will shorten (\ref{eqn:aj}) to
$\pi : a \rightarrow b$, or 
$ a \,
{\stackrel{\rm \pi}{\rightarrow}} \, b$.
The weight of $\pi$ is given by 
$w(\pi)=\prod_{j=1}^{m} w(e_j)$.
For any two vertices of $a, b \in V$,
the Green's function of walks 
$\{\pi : a \rightarrow b\}$
is defined by
\begin{equation}
W(a,b)=\sum_m \sum_{\pi: a \rightarrow b , |\pi|=m }
w(\pi).
\label{eqn:Green1}
\end{equation}
The matrix $W=(W(a,b))_{a, b \in V}$ is called
the {\it walk matrix} of the network $\Gamma$.

The loop-erased part of $\pi$,
denoted by $\LE(\pi)$, is defined recursively as follows.
If $\pi$ does not have self-intersections,
that is, all vertices $v_j, 0 \leq j \leq m$ are distinct, then
$\LE(\pi)=\pi$.
Otherwise, set $\LE(\pi)=\LE(\pi')$,
where $\pi'$ is obtained by removing the first loop
it makes.
In other words, if $(k, \ell), k < \ell$ is 
the smallest pair 
in the index set $\{j: 0 \leq j \leq m \}$
of $\{v_j\}_{j=0}^{m}$ in the sequence (\ref{eqn:aj}) such that
$v_k=v_{\ell}$, then the subsequence
$ v_k \,
{\stackrel{e_{k+1}}{\rightarrow}} \, 
v_{k+1} \, 
{\stackrel{e_{k+2}}{\rightarrow}} \,  
\cdots
{\stackrel{e_{\ell}}{\rightarrow}} \, 
v_{\ell} 
$
is removed from $\pi$ to obtain $\pi'$.

The loop-erasing operator $\LE$ maps
arbitrary walks to `walks without self-intersections',
which are usually called {\it self-avoiding walks} (SAWs).
Note that the map is many-to-one;
if $\zeta$ is a SAW obtained by applying $\LE$,
the set of inverse images
$\{\pi : \LE(\pi)= \zeta \}$
has more than one element in general.
For each SAW $\zeta$, the weight $\widetilde{w}(\zeta)$
is given by
\begin{equation}
\widetilde{w}(\zeta)=
\sum_{\pi : \LE(\pi)=\zeta}
w(\pi).
\label{eqn:LEweight}
\end{equation}
We consider the statistical ensemble of
SAWs with the weight (\ref{eqn:LEweight})
and call it {\it loop-erased random walks} (LERWs)
\cite{Ken00,LL10}.
Note that the LERW model is
different from the SAW model, 
in the sense that, though the configuration space of walks are the same, 
the weight of each walk is different
from each other.
In the SAW model we consider a statistical ensemble of
SAWs with the weight $\hat{w}(\zeta)=e^{-\beta |\zeta|}$,
where $e^{\beta}$ is the SAW connective constant,
while the weight of SAW in the LERW model
is given by the sum of weights
of all walks, which are the inverse images of the projection $\LE$
as shown by (\ref{eqn:LEweight}).

\begin{figure}
\includegraphics[width=0.6\linewidth]{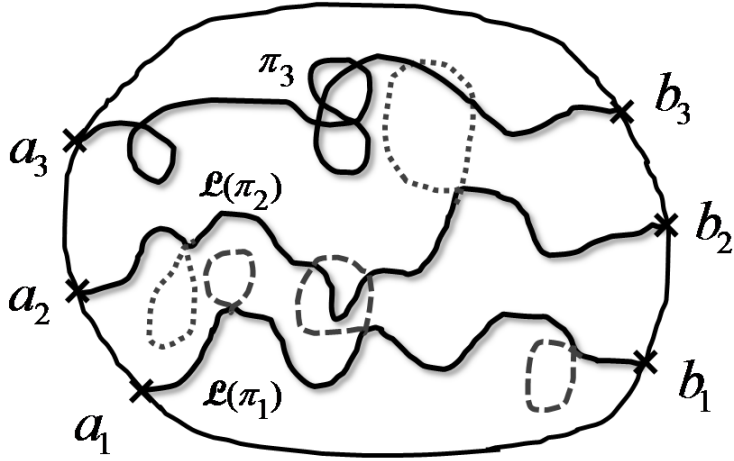}
\caption{The situation $\LE(\pi_j) \cap \pi_3=\emptyset,
j=1,2$ is illustrated in a planar domain $D$,
where $A=(a_1, a_2, a_3)$ and $B=(b_1, b_2, b_3)$
are all boundary points of $\partial D$.
In this figure, $\LE(\pi_1)$ and $\LE(\pi_2)$ 
denoted by solid curves are
the loop-erased parts of the walks
$\pi_1: a_1 \rightarrow b_1$ and
$\pi_2: a_2 \rightarrow b_2$, respectively,
where loops described by broken curves are erased.
The third walk $\pi_3: a_3 \rightarrow b_3$
can be self-intersecting, but it does not
intersect with $\LE(\pi_1)$ nor $\LE(\pi_2)$.
As a matter of course, $\LE(\pi_3)$ is a part of
$\pi_3$, and thus
$\LE(\pi_j) \cap \LE(\pi_3) = \emptyset, j=1,2$.
}
\label{fig:Fig1}
\end{figure}

Assume that $A=(a_1, a_2, \dots, a_N) \subset V$ and
$B=(b_1, b_2, \dots, b_N) \subset V$ are chosen
so that any walk from $a_j$ to $b_k$ intersects
any walk from $a_{j'}, j' > j$,
to $b_{k'}, k' < k$.
The weight of $N$-tuples of independent walks
$ a_1 \, {\stackrel{\rm \pi_1}{\rightarrow}} \, b_1, 
\dots, 
a_N \, {\stackrel{\rm \pi_N}{\rightarrow}} \, b_N $
is given by the product of $N$ weights
$\prod_{\ell=1}^N w(\pi_{\ell})$.
Then we consider $N$-tuples of walks
$(\pi_1, \pi_2, \dots, \pi_N)$ 
conditioned so that, for any
$1 \leq j < k \leq N$, the walk $\pi_k$
has no common vertices with the loop-erased part
of $\pi_j$;
\begin{equation}
\LE(\pi_j) \cap \pi_k=\emptyset, \quad
1 \leq j < k \leq N.
\label{eqn:LEW0}
\end{equation}
See Fig.\ref{fig:Fig1}.
By definition, $\LE(\pi_k)$ is a part of $\pi_k$, and thus
nonintersection of any pair of loop-erased parts
is concluded from (\ref{eqn:LEW0});
\begin{equation}
\LE(\pi_j) \cap \LE(\pi_k) = \emptyset,
\quad 1 \leq j < k \leq N.
\label{eqn:nonint}
\end{equation}

Fomin proved that total weight of $N$-tuples of walks
satisfying such a version of nonintersection condition
is given by the minor of
walk matrix,
$\det(W_{A,B}) \equiv
\det_{a \in A, b \in B}(W(a,b))$ \cite{For01}.
This minor is called {\it Fomin's determinant}
in the present paper and Fomin's formula is expressed by
the equality \cite{For01,KL05,LL10}
\begin{equation}
\det(W_{A,B}) = 
\sum_{\LE(\pi_j) \cap \pi_k=\emptyset, \,
j<k} 
\prod_{\ell=1}^{N} w(\pi_{\ell}).
\label{eqn:Fomin}
\end{equation}
It should be noted that the RHS of (\ref{eqn:Fomin})
seems to depend nontrivially on the
ordering of the sets $A$ and $B$,
but it is indeed invariant up to a sign
under the change of ordering,
since it is equal to the LHS of (\ref{eqn:Fomin}),
which is antisymmetric in exchanging any pair
of rows or columns \cite{For01}.

\section{Nonintersecting Systems of Loop-Erased Brownian Paths}

\subsection{Fomin's determinants for Brownian paths}

Random walks on a lattice can be regarded as
discrete approximations of a Brownian path.
For example, for a two-dimensional Brownian path $\gamma$
starting from a point $a$ and terminating at a point $b$
in a domain $D$ in the plane $\R^2$, we can consider a random walk
on a planar lattice $\Gamma$
embedded in $\R^2$ to approximate $\gamma$.
Assume that the lattice spacing of the planar lattice $\Gamma^{(n)}$ 
is $1/n, n \in \N$,
choose two vertices $a^{(n)}$ and $b^{(n)}$ such that
they are nearest vertices in $\Gamma^{(n)}$ to $a$ and $b$, respectively,
and denote the random walk on $\Gamma^{(n)}$
from $a^{(n)}$ to $b^{(n)}$ by $\pi^{(n)}$.
The continuum limit can be taken by getting $n \to \infty$
and $\pi^{(n)}$ will converge to a Brownian path 
$\gamma$ running from $a$ to $b$.
When the points $a$ and $b$ are on the boundary of a domain $D$,
and especially when the boundary is not smooth,
we need careful consideration for continuum limit
even for a single path.
See \cite{Koz06} for rigorous argument.

Then, if we can perform the same limiting procedure
$n \to \infty$ 
for Fomin's determinants of $N$-tuples of random walks
$\{\pi_1^{(n)}, \dots, \pi_N^{(n)}\}$ 
on planar lattices $\Gamma^{(n)}$, 
the limit will give the total weight of the $N$-tuples
of two-dimensional Brownian paths
$(\gamma_1, \dots, \gamma_N)$,
which satisfy the following condition,
\begin{equation}
\LE(\gamma_j) \cap \gamma_k= \emptyset,
\quad 1 \leq j < k \leq N,
\label{eqn:LEBP1}
\end{equation}
where $\LE(\gamma_j)$ is considered to be
the $n \to \infty$ limit of the sequence of loop-erased
parts $\{\LE(\pi_j^{(n)}) \}_{n \in \N}$
of discrete approximations 
$\{\pi_j^{(n)}\}_{n \in \N}$ of $\gamma_j$,
$1 \leq j \leq N$.

Recently Kozdron and Lawler \cite{KL05}
gave the mathematical justification of the above
mentioned continuum-limit-procedure
for simply connected planar domains $D$
in the complex plane $\C$,
where the initial and the final points
$A=\{a_j\}$ and $B=\{b_j\}$ of paths $\{\gamma_j\}$
can be put on the boundaries of the domains
$\partial D$. 
We should note that the characteristics of Brownian motion
look more similar to those of a surface than
those of a curve 
(see, for example, Chapter 1 of \cite{ID89}).
It implies that the Brownian path has loops on every scale
and then the loop-erasing procedure defined for random walks
in Sect.II does not make sense for Brownian motion,
since we can not decide which loop is the first one.
Kozdron and Lawler proved explicitly, however, that
the continuum limit of Fomin's determinant of the
Green's functions of random walks
converges to that of the Green's functions of 
Brownian motions \cite{KL05}.
This will enable us to discuss 
{\it nonintersecting systems} of 
LEBPs without dealing with
{\it individual} LEBP.
See a remark put at the end of the present paper
and \cite{KL05,KL07,Koz09} for more details.

One of the advantages of taking continuum limit
is the fact that the Green's function and
its normal derivatives at boundary points
(the Poisson kernel and the boundary Poisson kernel)
are obtained by solving the
Laplace equation with appropriate boundary conditions \cite{KL05}.
As a realization of the two-dimensional Brownian motion,
we consider a complex Brownian motion
$B_t=B_t^{({\rm R})}+\im B_t^{({\rm I})}$,
where $B_t^{({\rm R})}$ and $B_t^{({\rm I})}$ are independent
one-dimensional standard Brownian motions
satisfying $(dB_t^{({\rm R})})^2=(dB_t^{({\rm I})})^2=dt$
and $dB_t^{({\rm R})} dB_t^{({\rm I})}=0, t > 0$.
For a domain $D \subset \C$, let $p_{D}(t, a, b)$ be the
transition probability density of the complex Brownian motion
from $a \in D$ to $b \in D$ with duration $t \geq 0$
with the absorbing boundary condition at $\partial D$.
The Green's function for this Brownian motion is
defined by
\begin{equation}
G_D(a,b)
=\int_{0}^{\infty} p_{D}(t,a,b) dt.
\label{eqn:Green2}
\end{equation}
(Note that the summation with respect to
the length of walk $m=|\pi|$ in (\ref{eqn:Green1})
for the Green's function of random walks
is here replaced by the integral with respect
to duration of time of Brownian motion.) 
It is also the Green's function for the Poisson
equation 
$\Delta_a G_D(a,b)=\delta(a-b)$,
where $\Delta_a$ is the Laplacian with respect to
the variable $a$,
with the Dirichlet boundary condition 
$G_D(a,b)=0$ on $a \in \partial D$.
The complex Brownian motion is conformally
invariant in the sense that
the Green's function in a domain
$G_D(a,b), a, b \in D$ has the property
\begin{equation}
G_D(a, b)=G_{D'}(f(a), f(b))
\label{eqn:conformal}
\end{equation}
for any conformal transformation 
$f: D \to D'$.
Then we will select a suitable domain $D$ in $\C$
such that the Poisson equation can be analytically solved
and explicitly determine the Green's function
$G_D$, and then, though the equality (\ref{eqn:conformal}),
we can obtain $G_{D'}$ for other domain $D'$
by an appropriate conformal transformation.

For $a \in D$ and $b \in \partial D$, 
the {\it Poisson kernel} $H_{D}(a,b)$ is defined by
\begin{equation}
H_{D}(a,b)=\frac{1}{2} \lim_{\varepsilon \to 0}
\frac{1}{\varepsilon} G_D(a, b+\varepsilon \n_b),
\label{eqn:Poisson1}
\end{equation}where $\n_b$ denotes the inward
unit normal vector at $b \in \partial D$.
By this definition, we see that $H_D(a,b)$ solves 
the Laplace equation
$\Delta_a H_D(a,b)=0$ for $a \in D$, that is,
$H_D(a,b)$ is a {\it harmonic function} of $a \in D$,
and satisfies the boundary condition
\begin{equation}
\lim_{a \to \alpha: a \in D} H_D(a,b)
=\delta(\alpha-b)
\quad \mbox{for} \quad
\alpha \in \partial D.
\label{eqn:Poisson2}
\end{equation}
Moreover, we define the {\it boundary Poisson kernel}
$H_{\partial D}(a,b)$ for $a,b \in \partial D$ by
\begin{equation}
H_{\partial D}(a,b)=\lim_{\varepsilon \to 0}
\frac{1}{\varepsilon} H_D(a+\varepsilon \n_a, b).
\label{eqn:Poisson3}
\end{equation}
From the conformal invariance (\ref{eqn:conformal})
of the Green's function $G_D$, the following 
{\it conformal covariance} properties are derived;
if $f: D \to D'$ is a conformal transformation,
\begin{eqnarray}
\label{eqn:confcov1}
&& 
H_{D}(a,b)=|f'(b)| H_{D'}(f(a), f(b)),
\quad a \in D, b \in \partial D, \\
\label{eqn:confcov2}
&&
H_{\partial D}(a,b)=|f'(a)||f'(b)| H_{\partial D'}(f(a), f(b)),
\quad a, b \in \partial D.
\end{eqnarray}
See Section 2.6 of \cite{KL05} and
Chapters 2 and 5 of \cite{Law05}
for further information about Poisson kernels, 
boundary Poisson kernels and their conformal covariance.
We will study Fomin's determinants
of the form
$\det_{1 \leq j,k \leq N}[H_D(a_j, b_k)]$ 
and $\det_{1 \leq j, k \leq N}[H_{\partial D}(a_j, b_k)]$
in the following.

\subsection{System in a rectangular domain
and the crossing exponent}

\begin{figure}
\includegraphics[width=0.7\linewidth]{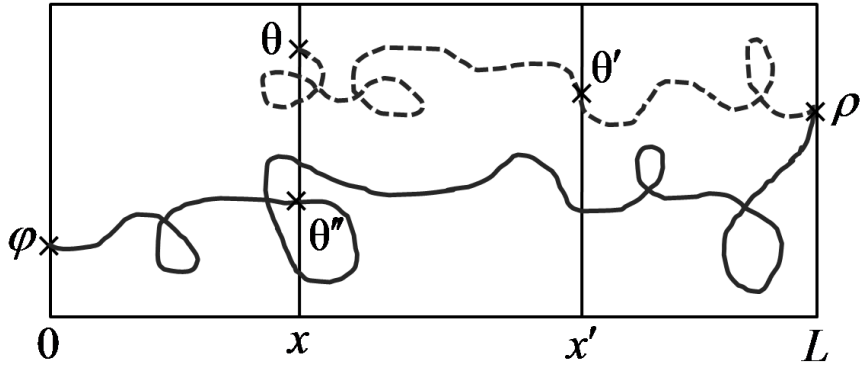}
\caption{The broken curve denotes a path of complex Brownian motion
starting from an inner point $x+\im \theta \in R_L$,
which makes first exit from $R_L$ at $L+\im \rho \in \partial R_L^{\rm R}$.
The solid curve does a path of complex Brownian motion, which enters $R_L$
at $\im \varphi \in R_L^{\rm L}$ and makes first exit from $R_L$
at $L+\im \rho \in \partial R_L^{\rm R}$.
The former path contributes to $H_{R_L}(x+\im \theta, L+\im \rho)$
and the latter does to $H_{\partial R_L}(\im \varphi, L+\im \rho)$,
respectively.
The point $x'+\im \theta'$ is the first passage point
on the line $\Re z=x'$ in $R_L$ of the former path,
and the point $x+\im \theta''$ is the first passage point
on the line $\Re z=x$ in $R_L$ of the latter path.
}
\label{fig:Fig2}
\end{figure}

In order to give explicit expressions for
Poisson kernel and boundary Poisson kernel, 
now we fix the domain $D$ as the
following rectangular domain
\begin{equation}
R_{L}=\Big\{ z=x+\im y \in \C :
0 < x < L, 0 < y < \pi \Big\}, 
\quad L >0.
\label{eqn:RL}
\end{equation}
As shown in Appendix A, for $0<x<L,
0 < \theta, \rho, \varphi < \pi$,
the Poisson kernel connecting an inner point
$x+\im \theta \in D$ and a point $L+\im \rho$
at the right boundary of $R_L$,
$\partial R_L^{\rm R}=\{L+\im y: 0 < y < \pi\}$, 
is given by
\begin{equation}
H_{R_{L}}(x+\im \theta, L+ \im \rho)
= \frac{2}{\pi} \sum_{n=1}^{\infty}
\frac{\sinh(n x) \sin(n \theta) \sin(n \rho)}
{\sinh(nL)},
\label{eqn:HRL}
\end{equation}
and the boundary Poisson kernel connecting
a boundary point $\im \varphi$ on the left boundary
of $R_L$,
$\partial R_L^{\rm L}=\{\im y : 0 < y < \pi\}$,
and a boundary point $L+ \im \rho$
on the right boundary $\partial R_L^{\rm R}$ is given by
\begin{equation}
H_{\partial R_L}(\im \varphi, L+\im \rho)
=\frac{2}{\pi} \sum_{n=1}^{\infty}
\frac{n \sin(n \varphi) \sin(n \rho)}
{\sinh(nL)}.
\label{eqn:HdRL}
\end{equation}
The definitions (\ref{eqn:Green2}), (\ref{eqn:Poisson1})
and (\ref{eqn:Poisson3}) imply that
$H_{R_L}(x+\im \theta, L+\im \rho) d \rho$
gives the total weight of the paths of complex Brownian motion
starting from the inner point $x+\im \theta \in R_L$,
which make first exit from the domain $R_L$ at a point
in the interval $[L+\im \rho, L+\im (\rho+d \rho)]$
on the right boundary $\partial R_L^{\rm R}$,
and that 
$H_{\partial R_L}(\im \varphi, L+\im \rho) d \varphi d \rho$
gives the total weight of the paths of complex Brownian motion,
which enter $R_L$ at a point in 
$[\im \varphi, \im(\varphi+d \varphi)]$ on $\partial R_L^{\rm L}$,
and make first exit from $R_L$ at a point in 
$[L+\im \rho, L+\im(\rho+d \rho)]$ on $\partial R_L^{\rm R}$.
(See Fig.\ref{fig:Fig2}.)

By the conservation of probability,
the following equalities should be satisfied;
for $0 < x < x' < L$,
\begin{eqnarray}
\label{eqn:CK1}
H_{R_L}(x+ \im \theta, L+\im \rho)
&=& \int_{0}^{\pi} d \theta' \,
H_{R_{x'}}(x+ \im \theta, x'+ \im \theta')
H_{R_L}(x'+ \im \theta', L+\im \rho), \\
\label{eqn:CK2}
H_{\partial R_{L}}(\im \varphi, L+ \im \rho)
&=& \int_{0}^{\pi} d \theta'' \,
H_{\partial R_{x}}(\im \varphi, x+ \im \theta'')
H_{R_L}(x+ \im \theta'', L+ \im \rho),
\end{eqnarray}
$0 < \varphi, \rho < \pi$.
As a matter of course, the validity of them can be
directly confirmed by applying the orthogonality relation of the
sine functions,
$\int_0^{\pi} d \theta \, \sin(n \theta) \sin (m \theta)
=(\pi/2) \delta_{n m}, n, m \in \N$,
to the expressions (\ref{eqn:HRL}) and (\ref{eqn:HdRL}).
We note that in Eq.(\ref{eqn:CK1}) the point $x'+\im \theta'$
is regarded as the {\it first passage point} on the line
$\Re z=x'$ in $R_L$ of the Brownian path
running from $x+\im \theta \in R_L$ to
$L+\im \rho \in \partial R_L^{\rm R}$,
and similarly that in Eq.(\ref{eqn:CK2}) $x+\im \theta''$
is regarded as the first passage point on the line $\Re z=x$
in $R_L$ of the Brownian path running from $\im \varphi \in 
\partial R_L^{\rm L}$ to $L+\im \rho \in \partial R_L^{\rm R}$.
See Fig.\ref{fig:Fig2}.

\begin{figure}
\includegraphics[width=0.7\linewidth]{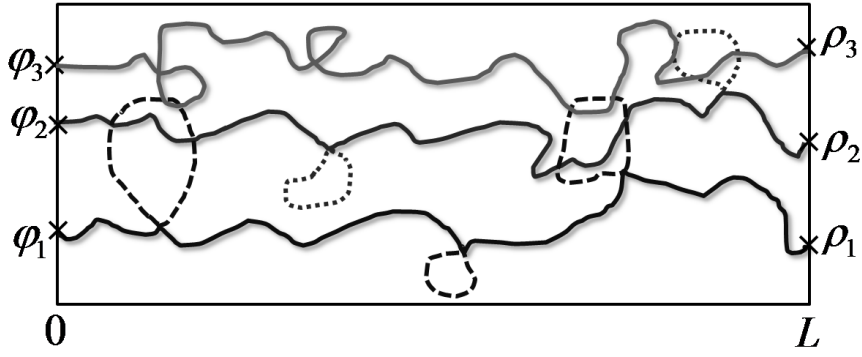}
\caption{The condition (\ref{eqn:LEBP1}) is illustrated
for $k=3, N=3$. 
The Brownian path $\gamma_3: \im \varphi_3 \rightarrow L+\im \rho_3$
in $R_L$ 
can be self-intersecting, but it does not
intersect with 
$\LE(\gamma_1): \im \varphi_1 \rightarrow L+\im \rho_1$ 
nor 
$\LE(\gamma_2): \im \varphi_2 \rightarrow L+\im \rho_2$.
By definition $\LE(\gamma_3)$ is a part of $\gamma_3$,
and thus $\LE(\gamma_1) \cap \LE(\gamma_3)=\emptyset$
and $\LE(\gamma_2) \cap \LE(\gamma_3)=\emptyset$.
}
\label{fig:Fig3}
\end{figure}

For $N \in \N$, let
$\W^{\pi}_N \equiv \{\vtheta=(\theta_1, \theta_2, \dots, \theta_N) :
0 < \theta_1 < \theta_2 < \cdots < \theta_N < \pi\}$.
Then, for $\vphi=(\varphi_1, \dots, \varphi_N) \in \W^{\pi}_N$,
$\vrho=(\rho_1, \dots, \rho_N) \in \W^{\pi}_N$,
consider Fomin's determinant
\begin{equation}
f^{\partial}_N(L, \vrho|\vphi)
\equiv \det_{1 \leq j, k \leq N} 
\Big[ H_{\partial R_L}
(\im \varphi_j, L+ \im \rho_k) \Big].
\label{eqn:fd1}
\end{equation}
If we write the path of $j$-th Brownian motion, $1 \leq j \leq N$, 
which enters $R_L$ at $\im \varphi_j \in \partial R_L^{\rm L}$ and
makes first exit from $R_L$ at $L+\im \rho_j \in
\partial R_L^{\rm R}$ as $\gamma_j$,
then (\ref{eqn:fd1}) gives the total weight
of the $N$-tuples of paths $(\gamma_1, \dots, \gamma_N)$
satisfying the nonintersection condition with
the loop-erased parts (\ref{eqn:LEBP1}).
We note again that it is a sufficient condition for
\begin{equation}
\LE(\gamma_j) \cap \LE(\gamma_k)=\emptyset,
\quad 1 \leq j< k \leq N.
\label{eqn:nonint2}
\end{equation}
See Fig.\ref{fig:Fig3}.

By multilinearlity of the determinant,
we find that Eq.(\ref{eqn:fd1}) with
Eq.(\ref{eqn:HdRL}) is written as
\begin{eqnarray} 
&& f^{\partial}_N(L, \vrho|\vphi) =
\left(\frac{2}{\pi}\right)^{N}
\sum_{\n=(n_1, \dots, n_N) \in {\bf N}^N} \prod_{j=1}^N
\frac{n_j}{\sinh(n_j L)} 
\nonumber\\
&& \qquad \qquad \times
\frac{1}{N!} \sum_{\sigma \in S_N}
\det_{1 \leq j,k \leq N}
\Big[ \sin(n_{\sigma(j)} \varphi_j)
\sin(n_{\sigma(j)} \rho_k) \Big] \nonumber\\
&& \quad =
\left(\frac{2}{\pi}\right)^{N}
\det_{1 \leq j, k \leq N} \Big[\sin(j \varphi_k) \Big]
\det_{1 \leq \ell, m \leq N} 
\Big[ \sin(\ell \rho_{m}) \big]
\times \sum_{\lambda}
a_{\lambda} \hat{s}_{\lambda}(\vphi)
\hat{s}_{\lambda}(\vrho), 
\label{eqn:fb2}
\end{eqnarray}
where $S_N$ is a set of all permutations $\{\sigma\}$
of $\{1,2, \dots, N\}$, $\lambda=(\lambda_1, \dots, \lambda_N)$
with $\lambda_j = n_{N-j+1}-(N-j+1), 1 \leq j \leq N$, and
\begin{eqnarray}
a_{\lambda} &=& \prod_{j=1}^{N}
\frac{\lambda_{j}+N-j+1}{\sinh \left((\lambda_j+N-j+1)L \right)},
\nonumber\\
\hat{s}_{\lambda}(\vphi) &=&
\frac{\displaystyle{
\det_{1 \leq j, k \leq N} \Big[
\sin\left( (\lambda_{k}+N-k+1) \varphi_{j} \right) \Big]}}
{\displaystyle{ \det_{1 \leq j, k \leq N} \Big[
\sin\left( (N-k+1) \varphi_{j} \right) \Big]}}.
\label{eqn:hats}
\end{eqnarray}
(This is a modified version of `Schur function expansion'
used in \cite{KT04,KT07}.)
We note that
\begin{equation}
\sin (\ell \theta)=
\sin \theta \left[
2^{\ell-1} (\cos \theta)^{\ell-1}
+ \sum_{s=1}^{[(\ell-1)/2]}
(-1)^{s} {\ell-s-1 \choose s} (2 \cos \theta)^{\ell-2s-1} \right],
\label{eqn:sin1}
\end{equation}
where, for $r \in \R$, 
$[r]$ denotes the greatest integer not greater than $r$.
Then for $\vtheta=(\theta_1, \dots, \theta_N) \in \W_N^{\pi}$
\begin{eqnarray}
\det_{1 \leq \ell, m \leq N} \Big[\sin(\ell \theta_m) \Big]
&=& \det_{1 \leq \ell, m \leq N}
\Big[ 2^{\ell-1} \sin \theta_{m} \Big\{
(\cos \theta_m)^{\ell-1} + {\cal O}((\cos \theta_m)^{\ell-3}) \Big\} \Big]
\nonumber\\
&=& 2^{N(N-1)/2} 
\hat{h}_N(\vtheta) 
\label{eqn:detsin}
\end{eqnarray}
with the function
\begin{equation}
\hat{h}_N(\vtheta)=
\prod_{j=1}^{N} \sin \theta_j
\prod_{1 \leq k < \ell \leq N} 
(\cos \theta_{\ell}-\cos \theta_k).
\label{eqn:tVande1}
\end{equation}
Since
$$
a_{\lambda} \simeq 2^{N} \prod_{j=1}^{N}
(\lambda_j +N-j+1) \,
e^{-L \sum_{j=1}^{N} (\lambda_j+N-j+1)}
\quad \mbox{as} \quad L \to \infty,
$$
and in particular for $\emptyset=(0,0, \dots, 0)$
$$
a_{\emptyset} \simeq
2^{N} N! \, e^{-LN(N+1)/2} 
\quad \mbox{as} \quad L \to \infty,
$$
we can conclude from the expansion (\ref{eqn:fb2}) that 
\begin{equation}
f^{\partial}_N(L, \vrho|\vphi)
\simeq 2^{N(N+1)} \pi^{-N} N! \, e^{-N(N+1)L/2}
\hat{h}_N(\vphi) \hat{h}_N(\vrho) \quad \mbox{as} \quad
L \to \infty. 
\label{eqn:asymfd}
\end{equation}
Since the simple $N$-product
of the boundary Poisson kernels
$\prod_{j=1}^{N} H_{\partial R_L}(\im \varphi_j, L+\im \rho_j)
\simeq \prod_{j=1}^{N} 2/\{\pi \sinh L \}
\simeq (4/\pi)^N e^{-NL}$ as $L \to \infty$, 
the ratio behaves
\begin{equation}
\Lambda_{R_{L}}(\vphi, \vrho)
\equiv \frac{f^{\partial}_N(L, \vrho|\vphi)}
{\prod_{j=1}^{N} H_{\partial R_L}(\im \varphi_j, L+\im \rho_j)}
\simeq c_N(\vphi, \vrho) e^{-\psi_N L}
\quad \mbox{as} \quad L \to \infty
\label{eqn:Lambda}
\end{equation}
with
\begin{equation}
\psi_N=\frac{1}{2} N(N-1), 
\label{eqn:cross}
\end{equation}
where 
$
c_N(\vphi, \vrho)
=2^{N(N-1)} N! 
\hat{h}_N(\vphi) \hat{h}_N(\vrho).
$
The exponent (\ref{eqn:cross}) is called
the {\it crossing exponent} \cite{KL05}.

\subsection{Probability density function for
first passage points of Brownian paths
underlying in nonintersecting LEBPs}

\begin{figure}
\includegraphics[width=0.5\linewidth]{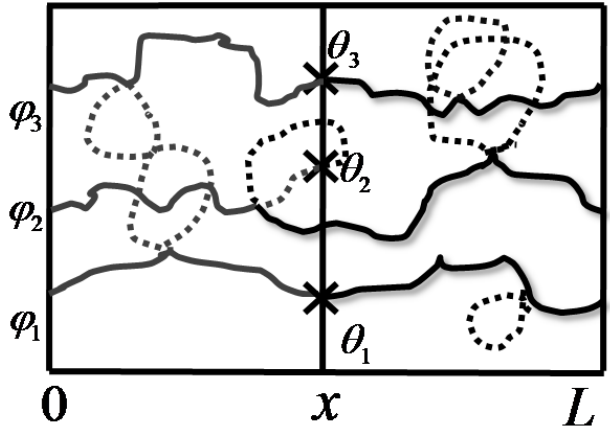}
\caption{Three Brownian paths $(\gamma_1, \gamma_2, \gamma_3)$
in $R_L=R_x \cup R_{\{x,L\}}$
starting from $(\im \varphi_1, \im \varphi_2, \im \varphi_3)$
and arriving at $\partial R_L^{\rm R}$
whose loop-erased parts
$(\LE(\gamma_1), \LE(\gamma_2), \LE(\gamma_3))$
are nonintersecting in the sense of Fomin.
The first passage points on the line $\Re z=x$
are $(x+\im \theta_1, x+\im \theta_2, x+\im \theta_3)$.
Since $x+\im \theta_2$ is in a loop of $\gamma_2$
as shown by a broken curve, it is {\it not}
included in $\LE(\gamma_2)$.
The probability density of points 
$(x+\im \theta_1, x+\im \theta_2, x+\im \theta_3)$
is given by Eq.(\ref{eqn:pL1}).
}
\label{fig:Fig4}
\end{figure}

We consider the integral of 
Fomin's determinant (\ref{eqn:fd1})
over all possible ordered sets of exits 
$\vrho \in \W^{\pi}_N$ at the right boundary $\partial R_L^{\rm R}$,
\begin{eqnarray}
{\cal N}^{\partial}_N(L, \vphi)
&\equiv& \int_{\W_N^{\pi}} d \vrho \,
f_N^{\partial}(L, \vrho|\vphi)
=
\int_{\W^{\pi}_N} d \vrho \,
\det_{1 \leq j, k \leq N} \Big[
H_{\partial R_L}(\im \varphi_j, L+\im \rho_k) \Big]
\nonumber\\
\label{eqn:N1}
&=& \int_{\W^{\pi}_N} d \vrho \,
\det_{1 \leq j, k \leq N} \left[
\int_{0}^{\pi} d \theta \,
H_{\partial R_x}(\im \varphi_j, x+ \im \theta)
H_{R_L}(x+ \im \theta, L+ \im \rho_k) \right],
\end{eqnarray}
where $d \vrho=\prod_{j=1}^{N} d \rho_j$,
$0 < x < L$, 
and Eq.(\ref{eqn:CK2}) has been used
in the last equality.
Applying the Heine identity
\begin{equation}
\int d \z \det_{1 \leq j,k \leq N}
\Big[\phi_j(z_k) \Big] \det_{1 \leq \ell,m \leq N}
\Big[\psi_\ell(z_m) \big] = \det_{1 \leq j,k \leq N}
\Big[\int d z \, \phi_j(z) \psi_k(z) \Big],
\label{eqn:Heine}
\end{equation}
which is valid for square integrable functions
$\phi_j, \psi_j, 1 \leq j \leq N$, 
it is written as
\begin{eqnarray}
&& {\cal N}^{\partial}_N(L, \vphi)
= \int_{\W^{\pi}_N} d \vrho 
\int_{\W^{\pi}_N} d \vtheta \,
\det_{1 \leq j, k \leq N}
\Big[ H_{\partial R_x}(\im \varphi_j, x+ \im \theta_k) \Big]
\det_{1 \leq \ell, m \leq N}
\Big[ H_{R_{L}}(x+\im \theta_{\ell}, L+\im \rho_{m}) \Big]
\nonumber\\
&& \qquad =
\int_{\W^{\pi}_N} d \vtheta \,
f^{\partial}_N(x, \vtheta|\vphi)
\int_{\W^{\pi}_N} d \vrho \,
\det_{1 \leq \ell, m \leq N}
\Big[ H_{R_{L}}(x+ \im \theta_{\ell}, L+ \im \rho_{m}) \Big].
\label{eqn:N2}
\end{eqnarray}
Then, if we introduce the integral
\begin{equation}
{\cal N}_N(x, L, \vtheta)
= \int_{\W^{\pi}_N} d \vrho 
f_N(x, \vtheta; L, \vrho), \quad 0 < x < L,
\label{eqn:NML1}
\end{equation}
of Fomin's determinant for the Poisson kernels
\begin{equation}
f_N(x, \vtheta; L, \vrho)
= \det_{1 \leq j, k \leq N}
\Big[ H_{R_{L}}(x+ \im \theta_j, L+ \im \rho_k) \Big],
\label{eqn:fN}
\end{equation}
$\vtheta, \vrho \in \W_N^{\pi}$, 
and divide the both sides of (\ref{eqn:N2}) 
by ${\cal N}^{\partial}_N(L, \vphi)$, we 
obtain the equality
\begin{equation}
1=\int_{\W^{\pi}_N} d \vtheta \,
f^{\partial}_N(x, \vtheta|\vphi)
\frac{{\cal N}_N(x, L, \vtheta)}
{{\cal N}^{\partial}_N(L, \vphi)}.
\label{eqn:norm}
\end{equation}
Then, given $\vphi \in \W_N^{\pi}$, if we put
\begin{equation}
p^L_N(x, \vtheta|\vphi)
= f^{\partial}_N(x, \vtheta|\vphi)
\frac{{\cal N}_N(x, L, \vtheta)}
{{\cal N}^{\partial}_N(L, \vphi)},
\quad 0 < x < L, 
\label{eqn:pL1}
\end{equation}
for $\vtheta \in \W_N^{\pi}$, 
it can be regarded as the probability density function.
As illustrated in Fig.\ref{fig:Fig4},
here we consider the rectangular domain (\ref{eqn:RL})
as a union of the two rectangular domains
$R_x=\{z \in \C: 0 < \Re z < x, 0 < \Im z < \pi\}$
and $R_{\{x, L\}}=\{z \in \C: x \leq \Re z < L,
0 < \Im z < \pi\}$,
which are adjacent to each other at a line
$\Re z=x$.
We have considered $N$-tuples of Brownian paths
$(\gamma_1, \dots, \gamma_N)$
starting from $(\im \varphi_1, \dots, \im \varphi_N)$,
all of which run inside of the domain
$R_L=R_x \cup R_{\{x, L\}}$ until
arriving at the right boundary $\partial R_L^{\rm R}$,
under the condition that $(\LE(\gamma_1), \dots, \LE(\gamma_N))$
makes a system of nonintersecting LEBPs in the sense of
Fomin (\ref{eqn:LEBP1}).
In an ensemble of such $N$-tuples of Brownian paths,
Eq.(\ref{eqn:pL1}) gives the
probability density of the event such that
the first passage points of $(\gamma_1, \dots, \gamma_N)$
on the line $\Re z=x$ are 
$(x+\im \theta_1, \dots, x+\im \theta_N)$.
That is, for $1 \leq j \leq N$, 
$\gamma_j$ makes a first exit from the left 
rectangular domain $R_x$ and enters the right 
rectangular domain $R_{\{x, L\}}$ at $x+\im \theta_j$.
The path $\gamma_j$, which made a first exit from $R_x$
at $x+\im \theta_j$, can reenter $R_x$
at different point $x+\im \theta'$ on the line $\Re z=x$.
As shown by the path $\gamma_2$ starting from $\im \varphi_2$
in Fig.\ref{fig:Fig4}, the path $\gamma_j$
can make a loop passing the line $\Re z=x$,
and if the first passage point $x +\im \theta_j$
is included in such a loop, that point can {\it not} be included
in $\LE(\gamma_j)$.
In other words, Eq.(\ref{eqn:pL1}) gives 
the probability density for 
$N$ nonintersecting loop-erased Brownian paths
$(\LE(\gamma_1), \dots, \LE(\gamma_N))$ 
starting from the points
$(\im \varphi_1, \dots, \im \varphi_N)$ and 
satisfying the condition (\ref{eqn:LEBP1})
in $R_L$, such that the underlying Brownian paths 
$(\gamma_1, \dots, \gamma_N)$
are realized so that they
first arrive at the vertical line $\Re z=x < L$
at $(x+\im \theta_1, \dots, x+ \im \theta_N)$.

Now we consider the system
in the limit $L \to \infty$.
By applying the asymptotics (\ref{eqn:asymfd}),
we have
\begin{eqnarray}
{\cal N}^{\partial}_N(L, \vphi)
&=& \int_{\W^{\pi}_N}
d \vrho \, 
f^{\partial}_N(L, \vrho| \vphi) \nonumber\\
&\simeq& 2^{N(N+1)} \pi^{-N} N! \, e^{-N(N+1)L/2}
\hat{h}_{N}(\vphi)
\times \int_{\W^{\pi}_N}
d \vrho \,
\hat{h}_{N}(\vrho),
\label{eqn:AsymNd}
\end{eqnarray}
as $L \to \infty$.
Similarly, we can obtain the following asymptotics
of (\ref{eqn:NML1})
\begin{eqnarray}
&& {\cal N}_N(x ,L, \vtheta)
= \int_{\W^{\pi}_N}
d \vrho \,
\det_{1 \leq j, k \leq N} 
\left[ \frac{2}{\pi} \sum_{n=1}^{\infty}
\frac{\sinh(n x) \sin (n \theta_j) \sin (n \rho_k)}
{\sinh(nL)} \right]
\nonumber\\
&& \quad \simeq 2^{N(N+1)} \pi^{-N}
e^{-N(N+1)L/2} 
\hat{h}_N(\vtheta) 
\prod_{j=1}^{N} \sinh(j x) \int_{\W^{\pi}_N}
d \vrho \,\hat{h}_{N}(\vrho),
\label{eqn:AsymN}
\end{eqnarray}
as $L \to \infty$.
Then, given $\vphi \in \W_N^{\pi}$,
for $\vtheta \in \W_N^{\pi}$, 
\begin{eqnarray}
p_N(x,\vtheta|\vphi) &\equiv&
\lim_{L \to \infty} p^{L}_N(x, \vtheta|\vphi)
\nonumber\\
&=& C_{N}(x)
f^{\partial}_N(x, \vtheta|\vphi)
\frac{
\hat{h}_{N}(\vtheta)}
{\hat{h}_{N}(\vphi)},
\quad 0 < x < \infty,
\label{eqn:pN1}
\end{eqnarray}
where
\begin{equation}
C_{N}(x)
= \frac{1}{N!} \prod_{j=1}^{N} \sinh(j x).
\label{eqn:CNM}
\end{equation}

\subsection{Joint distribution
of first passage points in a sequence of chambers}

Let $M \in \N$ and 
$0 < x_1 < x_2 < \dots < x_M < L < \infty$.
Here we consider $M$ vertical lines in $R_N$
at $\Re z=x_m, 1 \leq m \leq M$.
For $(x_m, x_{m+1})$,
$\vtheta^{(m)}=(\theta^{(m)}_1, \dots, \theta^{(m)}_N)
\in \W^{\pi}_N$,
$\vtheta^{(m+1)}=(\theta^{(m+1)}_1, \dots, \theta^{(m+1)}_N)
\in \W^{\pi}_N$, $1 \leq m \leq M-1$, 
we define
\begin{equation}
q^L_N(x_m, \vtheta^{(m)}; x_{m+1}, \vtheta^{(m+1)})
=\frac{{\cal N}_N(x_{m+1}, L, \vtheta^{(m+1)})}
{{\cal N}_N(x_m, L, \vtheta^{(m)})}
f_N(x_m, \vtheta^{(m)}; x_{m+1}, \vtheta^{(m+1)}).
\label{eqn:qL}
\end{equation}
Then by definition (\ref{eqn:pL1}), we have
\begin{eqnarray}
&& \int_{\W^{\pi}_N} d \vtheta^{(1)} \,
p^L_N(x_1, \vtheta^{(1)}|\vphi)
q^L_N(x_1, \vtheta^{(1)}; x_2, \vtheta^{(2)})
\nonumber\\
&=& \int_{\W^{\pi}_N} d \vtheta^{(1)}
f^{\partial}_N(x_1, \vtheta^{(1)}|\vphi)
\frac{{\cal N}_N(x_1, L, \vtheta^{(1)})}
{{\cal N}^{\partial}_N(L, \vphi)}
\frac{{\cal N}_N(x_2, L, \vtheta^{(2)})}
{{\cal N}_N(x_1, L, \vtheta^{(1)})}
f_N(x_1, \vtheta^{(1)}; x_2, \vtheta^{(2)})
\nonumber\\
&=& \frac{{\cal N}_N(x_2, L, \vtheta^{(2)})}
{{\cal N}^{\partial}_N(L, \vphi)}
\int_{\W^{\pi}_N} d \vtheta^{(1)} \,
f^{\partial}_N(x_1, \vtheta^{(1)}|\vphi)
f_N(x_1, \vtheta^{(1)}; x_2, \vtheta^{(2)}).
\nonumber
\end{eqnarray}
By the Heine identity (\ref{eqn:Heine}) and 
the equality (\ref{eqn:CK2}), 
\begin{eqnarray}
&& \int_{\W^{\pi}_N} d \vtheta^{(1)} \,
f^{\partial}_N(x_1, \vtheta^{(1)}|\vphi)
f_N(x_1, \vtheta^{(1)}; x_2, \vtheta^{(2)})
\nonumber\\
&=& \int_{\W^{\pi}_N} d \vtheta^{(1)} 
\det_{1 \leq j, k \leq N}
\Big[ H_{\partial R_{x_1}}(\im \varphi_j, x_1+\im \theta^{(1)}_k) \Big]
\det_{1 \leq \ell, m \leq N} \Big[
H_{R_{x_2}}(x_1+\im \theta^{(1)}_{\ell},
x_2+\im \theta^{(2)}_m) \Big]
\nonumber\\
&=& \det_{1 \leq j, k \leq N}
\left[ \int_0^{\pi} d \theta^{(1)} \,
H_{\partial R_{x_1}}(\im \varphi_j, x_1+\im \theta^{(1)})
H_{R_{x_2}}(x_1+\im \theta^{(1)}, x_2+\im \theta^{(2)}_k) \right]
\nonumber\\
&=& \det_{1 \leq j, k \leq N}
\Big[ H_{\partial R_{x_2}}(\im \varphi_j, 
x_2+ \im \theta^{(2)}_k) \Big]
=f^{\partial}_N(x_2, \vtheta^{(2)}|\vphi),
\nonumber
\end{eqnarray}
and thus we obtain the equality
$$
p^{L}_N(x_2, \vtheta^{(2)}|\vphi)
=\int_{\W^{\pi}_N} d \vtheta^{(1)} \,
p^L_N(x_1, \vtheta^{(1)}|\vphi)
q^L_N(x_1, \vtheta^{(1)}; x_2, \vtheta^{(2)}).
$$
Obviously this equality can be generalized as
\begin{equation}
p^{L}_N(x_{m+1}, \vtheta^{(m+1)}|\vphi)
=\int_{\W^{\pi}_N} d \vtheta^{(m)} \,
p^L_N(x_m, \vtheta^{(m)}|\vphi)
q^L_N(x_m, \vtheta^{(m)}; x_{m+1}, \vtheta^{(m+1)})
\label{eqn:pita}
\end{equation}
for $1 \leq m \leq M-1$.

\begin{figure}
\includegraphics[width=1.0\linewidth]{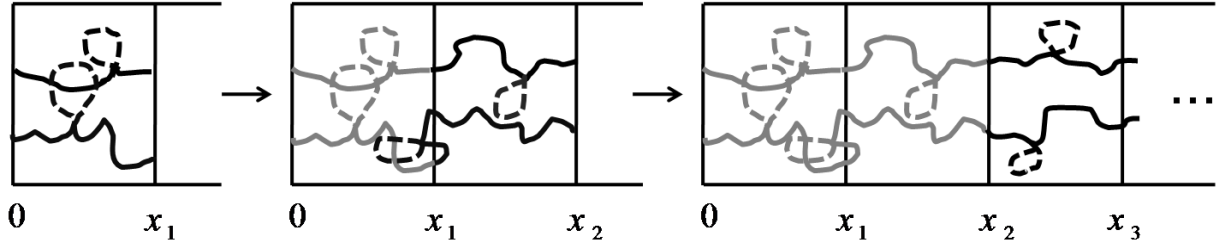}
\caption{Eq.(\ref{eqn:jointpL}) gives the probability
density function for joint distributions of $M$
sets of first passage points on the lines
$\Re z=x_m, 1 \leq m \leq M$, of
$N$-tuples of Brownian paths underlying in the
nonintersecting LEBPs.
}
\label{fig:Fig5}
\end{figure}

Then, given $\vphi \in \W_N^{\pi}$, 
if we introduce a function of
$\vtheta^{(m)}=(\theta^{(m)}_1, \dots, \theta^{(m)}_N) \in \W_N^{\pi},
1 \leq m \leq M$ by
\begin{eqnarray}
&& p^L_N(x_1, \vtheta^{(1)}; x_2, \vtheta^{(2)}; \dots ;
x_M, \vtheta^{(M)}|\vphi)
\nonumber\\
&& = p^L_N(x_1, \vtheta^{(1)}|\vphi)
\prod_{m=1}^{M-1} q^L_N(x_m, \vtheta^{(m)}; x_{m+1}, \vtheta^{(m+1)})
\nonumber\\
&& = \frac{{\cal N}_N(x_M, L, \vtheta^{(M)})}
{{\cal N}^{\partial}_N(L, \vphi)}
f^{\partial}_N(x_1, \vtheta^{(1)}|\vphi)
\prod_{m=1}^{M-1} f_N(x_m, \vtheta^{(m)}; x_{m+1},
\vtheta^{(m+1)}),
\label{eqn:jointpL}
\end{eqnarray}
it can be regarded as the probability density function,
since it is well normalized as
$$
\prod_{m=1}^{M} \int_{\W^{\pi}_N} d \vtheta^{(m)}
p^L_N(x_1, \vtheta^{(1)}; \dots ;
x_M, \vtheta^{(M)} | \vphi)
\nonumber\\
= \int_{\W^{\pi}_N} d \vtheta^{(M)} \,
p^L_N(x_M, \vtheta^{(M)}|\vphi)=1
$$
by (\ref{eqn:pita}) and (\ref{eqn:norm}).
As shown by Fig.\ref{fig:Fig5},
here we think that
$R_L$ has $M+1$ `chambers'
$R_{\{x_{m-1}, x_m\}}, 1 \leq m \leq M+1$;
$R_L=\bigcup_{m=1}^{M+1} R_{\{x_{m-1}, x_m\}}$
with $x_0 \equiv 0, x_{M+1}=L$,
where $R_{\{x_{m-1}, x_m\}}$ and
$R_{\{x_m, x_{m+1}\}}$ are adjacent to each other
at the line $\Re z=x_m, 1 \leq m \leq M$.
Under the initial condition that $\gamma_j$ starts from $\im \varphi_j$,
$1 \leq j \leq N$ and the nonintersection condition
of $(\LE(\gamma_1), \dots, \LE(\gamma_N))$ in the sense of
Fomin (\ref{eqn:LEBP1}),
Eq.(\ref{eqn:jointpL}) gives the probability density
function for joint distributions
of the first passage points 
$(x_m+\im \theta^{(m)}_1, \dots, x_m+\im \theta^{(m)}_N)$
at which $\gamma_j$'s first pass from the $m$-th
chamber $R_{\{x_{m-1}, x_m\}}$
to the $(m+1)$-th chamber $R_{\{x_m, x_{m+1}\}}$
on the line $\Re z=x_m, 1 \leq m \leq M$.

By the asymptotics (\ref{eqn:AsymNd}) and (\ref{eqn:AsymN}), 
the limit $L \to \infty$ is taken as
\begin{eqnarray}
&& p_N(x_1, \vtheta^{(1)}; \dots ;
x_M, \vtheta^{(M)} | \vphi)
\nonumber\\
&& \equiv \lim_{L \to \infty}
p^L_N(x_1, \vtheta^{(1)}; \dots ;
x_M, \vtheta^{(M)} | \vphi)
\nonumber\\
&& = C_N(x_M) \frac{\hat{h}_N(\vtheta^{(M)})}
{\hat{h}_N(\vphi)}
f^{\partial}_N(x_1, \vtheta^{(1)}|\vphi)
\prod_{m=1}^{M-1} f_N(x_m, \vtheta^{(m)};
x_{m+1}, \vtheta^{(m+1)}),
\label{eqn:jointp}
\end{eqnarray}
$\vtheta^{(m)} \in \W_N^{\pi}, 1 \leq m \leq M$,
given the initial condition $\vphi \in \W_N^{\pi}$.

\section{Determinantal Correlation Functions}

\subsection{Special initial condition}

Let $A_{\lambda}(\vphi)$ be the numerator of
$\hat{s}_{\lambda}(\vphi)$ given by (\ref{eqn:hats}).
By the expansion formula (\ref{eqn:sin1}), we find
\begin{eqnarray}
&& A_{\lambda}(\vphi) 
= \det_{1 \leq j, k \leq N}
\Big[ \sin \varphi_{j} \Big\{
2^{\lambda_{k}+N-k} (\cos \varphi_{j})^{\lambda_{k}+N-k}
+ {\cal O}((\cos \varphi_{j})^{\lambda_{k}+N-k-2}) \Big\} \Big]
\nonumber\\
&& \quad = 2^{\sum_{k=1}^{N}(\lambda_{k}+N-k)}
\prod_{j=1}^{N} \sin \varphi_{j} 
\det_{1 \leq j, k \leq N}
\Big[ (\cos \varphi_j)^{\lambda_{k}+N-k}+ 
{\cal O}((\cos \varphi_{j})^{\lambda_{k}+N-k-2})
\Big] \nonumber\\
&& \quad = 2^{\sum_{k=1}^{N}(\lambda_{k}+N-k)}
\hat{h}_N(\vphi) \tilde{A}_{\lambda}(\vphi),
\label{eqn:tA}
\end{eqnarray}
where $\tilde{A}_{\lambda}(\vphi)$ is a symmetric function
of $\{\cos \varphi_j\}_{j=1}^{N}$ with
degree $\sum_{k=1}^{N} \lambda_k$.
If we write $(\pi/2, \dots, \pi/2)$ as $\vpi/2$,
$\lim_{\vphi \to \vpi/2} \tilde{A}_{\lambda}(\vphi)
=\delta_{\lambda, \emptyset}$.
Then (\ref{eqn:fb2}) and (\ref{eqn:detsin}) give
\begin{equation}
\lim_{\vphi \to \vpi/2} 
\frac{f^{\partial}_N(L, \vrho|\vphi)}
{\hat{h}_N(\vphi)}
=\frac{2^{N^2}}{\pi^N C_N(L)} \hat{h}_N(\vrho)
\label{eqn:pi/2_1}
\end{equation}
with the factor (\ref{eqn:CNM}).

Applying (\ref{eqn:pi/2_1}) to (\ref{eqn:pN1}),
we obtain the limit distribution
\begin{equation}
p_N^{\pi/2}(x, \vtheta) \equiv \lim_{\vphi \to \vpi/2} 
p_N(x, \vtheta|\vphi)
= \frac{2^{N^2}}{\pi^N} 
(\hat{h}_N(\vtheta))^2.
\label{eqn:pN0}
\end{equation}
The fact that it is well normalized, {\it i.e.}
$\int_{\W^{\pi/2}_N} p_N^{\pi/2}(\vtheta) d \vtheta=1$,
is directly confirmed by using the $\gamma=1$ case
of `the Tchebichev version' of the Selberg integral
\begin{eqnarray}
&& \int_{-1}^{1} \cdots \int_{-1}^{1}
\left|\prod_{1 \leq \ell < m \leq N}
(\xi_m-\xi_{\ell}) \right|^{2\gamma}
\prod_{j=1}^{N}(1-\xi_j^2)^{1/2} d\xi_j
\nonumber\\
&& \qquad =2^{\gamma N(N-1)+2N}
\prod_{j=0}^{N-1} \frac{
\Gamma(1+\gamma+j\gamma)(\Gamma(\gamma j+3/2))^2}
{\Gamma(1+\gamma) \Gamma(\gamma(N+j-1)+3)},
\nonumber
\end{eqnarray}
which is given as Eq.(17.6.4) in \cite{Meh04}.
Under this special initial condition $\vpi/2$,
the joint distribution function (\ref{eqn:jointp})
for $L \to \infty$ becomes
\begin{eqnarray}
&& p_N^{\pi/2}(x_1, \vtheta^{(1)}; x_2, \vtheta^{(2)};
\dots; x_M, \vtheta^{(M)})
\nonumber\\
&& \quad \equiv \lim_{\vphi \to \vpi/2}
p_N(x_1, \vtheta^{(1)}; x_2, \vtheta^{(2)};
\dots; x_M, \vtheta^{(M)}|\vphi)
\nonumber\\
&& \quad =
\frac{2^{N^2}}{\pi^N}
\frac{C_N(x_M)}{C_N(x_1)}
\hat{h}_N(\vtheta^{(1)})
\prod_{m=1}^{M-1}
f_N(x_m, \vtheta^{(m)}; x_{m+1}, \vtheta^{(m+1)})
\hat{h}_N(\vtheta^{(M)})
\label{eqn:pNjoint0}
\end{eqnarray}
for any $M \in \N, 0 < x_1 < \cdots < x_M < \infty$.
As a matter of course, if we set $M=1$,
(\ref{eqn:pNjoint0}) is reduced to be (\ref{eqn:pN0}).

\subsection{Multiple correlation functions}

For $\vtheta^{(m)} \in \W^{\pi}_N$, 
${N'}_m \in \{1,2, \dots, N\}, 1 \leq m \leq M$,
we put
$\vtheta^{(m)}_{N'}=(\theta^{(m)}_1, \dots, \theta^{(m)}_{N'}),
1 \leq m \leq M$.
For a sequence $\{N_m\}_{m=1}^{M}$ of positive
integers less than or equal to $N$, we define the
$(N_1, \dots, N_M)$-multiple correlation function
by
\begin{eqnarray}
&& \rho_N^{\pi/2}(x_1, \vtheta^{(1)}_{N_1};
\cdots ; x_M, \vtheta^{(M)}_{N_M})
\nonumber\\
&& =
\prod_{m=1}^{M} \prod_{j=N_{m+1}}^{N}
\int_0^{\pi} d \theta^{(m)}_j \,
p_N^{\pi/2}(x_1, \vtheta^{(1)}; \dots ; x_M, \vtheta^{(M)})
\prod_{m=1}^{M} \frac{1}{(N-N_m)!}.
\label{eqn:rho1}
\end{eqnarray}

For $0< x < \infty, 0< \theta < \pi$, 
we introduce two systems of functions as
\begin{eqnarray}
&& \phi_n(x,\theta) = \sqrt{\frac{2}{\pi}} 
\frac{\sin(n\theta)}{\sinh(nx)}, \nonumber\\
&& 
\hat{\phi}_n(x,\theta) = \sqrt{\frac{2}{\pi}}
\sinh(nx) \sin(n \theta),
\quad n \in \N.
\label{eqn:phi1}
\end{eqnarray}
It is easy to confirm the equalities
\begin{eqnarray}
&& \int_0^{\pi} \phi_n(x,\theta)
H_{R_{x'}}(x+\im \theta, x'+ \im \theta') d \theta
= \phi_n(x',\theta'), 
\quad 0 < \theta' < \pi, 
\nonumber\\
&&
\int_0^{\pi} 
H_{R_{x'}}(x+ \im \theta, x'+ \im \theta')
\hat{\phi}_n(x', \theta') d \theta' 
= \hat{\phi}_n(x, \theta),
\quad 0 < \theta < \pi, 
\label{eqn:phi2}
\end{eqnarray}
$0 < x < x' < \infty, n \in \N$.
By using them, the probability density function of joint distributions
(\ref{eqn:pNjoint0}) is rewritten as follows,
\begin{eqnarray}
&& p_N^{\pi/2}(x_1,\vtheta^{(1)}; \cdots ;x_M,\vtheta^{(M)}) 
\nonumber\\ 
&& \qquad =
\det_{1 \leq j,k \leq N} [\phi_j(x_1,\theta_k^{(1)})]
\prod_{m=1}^{M-1} \det_{1 \leq \alpha,\beta \leq N}
[H_{R_{x_{m+1}}} (x_m+ \im \theta_{\alpha}^{(m)},
x_{m+1}+ \im \theta_{\beta}^{(m+1)})]
\nonumber\\
&& \qquad \qquad \times
\det_{1 \leq p, q \leq N} 
[\hat{\phi}_p(x_M, \theta_q^{(M)})].
\label{eqn:pNjoint0B}
\end{eqnarray}
We have found that this product form of determinants
is exactly the same as Eq.(4.5) in \cite{KT07}
given for the multitime distribution function
of the noncolliding Brownian motion.
Then following the argument given in Section 4 of 
\cite{KT07}, the following result is obtained.

\vskip 0.5cm
\noindent{\bf Theorem 1.} \quad
Any multiple correlation function (\ref{eqn:rho1})
is given by a determinant
\begin{equation}
\rho_N^{\pi/2}(x_1,\vtheta_{N_1}^{(1)};x_2,\vtheta_{N_2}^{(2)};
\cdots ;x_M,\vtheta_{N_M}^{(M)}) = 
\det_{1 \leq j \leq N_m, 1 \leq k \leq N_n, 
1 \leq m, n \leq M}
\Big[ \K_N^{\pi/2}(x_m, \theta^{(m)}_j;
x_n, \theta^{(n)}_k) \Big]
\label{eqn:Determinantal1}
\end{equation}
with the correlation kernel
\begin{eqnarray}
&& \K_N^{\pi/2}(x,\theta; x',\theta') =
\sum_{n=1}^{N} \phi_n(x,\theta) {\hat \phi}_n(x',\theta') 
\nonumber\\
\label{eqn:K1}
&& \qquad = \frac{2}{\pi} \sum_{n=1}^N 
\frac{\sinh(n x')}{\sinh(n x)}
\sin(n\theta) \sin(n\theta'), \quad \mbox{if} \quad x \leq x', 
\\
&& \K_N^{\pi/2}(x,\theta; x',\theta') =
\sum_{n=1}^{N} \phi_n(x,\theta) {\hat \phi}_n(x',\theta')
- H_{R_x} (x'+ \im \theta', x+ \im \theta) 
\nonumber\\
\label{eqn:K2}
&& \qquad= -\frac{2}{\pi} \sum_{n=N+1}^{\infty}
\frac{\sinh(n x')}{\sinh(n x)}
\sin(n\theta) \sin(n\theta'), \quad \mbox{if} \quad x >x',
\end{eqnarray}
$0< x, x' < \infty, 0 < \theta, \theta' < \pi$.
\vskip 0.5cm

By using the terminology of probability theory,
we can say that the ensemble of first passage points
$\{x_m + \im \theta^{(m)}_{j} :1 \leq j \leq N, 1 \leq m \leq M\}$ 
is a determinantal point process
(or a Fermion point process) \cite{Sos00,ST03,HKPV09}.
Note that the correlation kernel 
given by (\ref{eqn:K1}) and (\ref{eqn:K2}) 
is asymmetric in the ordering of $x$ and $x'$;
$\K_N^{\pi/2}(x, \theta; x', \theta') \not=
\K_N^{\pi/2}(x', \theta'; x, \theta)$
for $x \not=x'$.
Such kind of asymmetric correlation kernel was
first derived by Eynard and Mehta 
for two-matrix models in random matrix theory
\cite{EM98,Meh04}. See \cite{BR05,Forr10,KT10c} for recent
study on the Eynard-Mehta type determinantal
correlations.

\section{Conformal Transformation to Other Domain}

In the previous section, we considered a nonintersecting 
system of LEBPs in the half-infinite-strip domain,
which is divided into $M$ rectangular chambers
and one half-infinite strip by $M$ straight lines
on $\Re z=x_m, 1 < x_1 < \cdots < x_M < \infty$.
For the underlying system of $N$-tuples of Brownian paths,
whose loop-erased parts give the nonintersecting LEBPs,
Theorem 1 gives the determinantal correlation functions
for first passage points on the lines 
$\Re z=x_m, 1 \leq m \leq M$.
In order to demonstrate that the result can be
conformally mapped to other domain consisting of
a sequence of chambers in different shapes,
here we show a conformal transformation
by an entire function $w=f(z)=e^z$.

By this conformal transformation, 
the half-infinite-strip domain
$R=\{z \in \R: \Re z >0, 0 < \Im z < \pi\}$
is mapped to the domain
$\Omega=\{z = r e^{\im \theta} \in \C :
r>1, 0 < \theta < \pi\}$.
The rectangular chambers $R_{\{x_{m-1}, x_m\}}, 1 \leq m \leq M$,
are mapped to the chambers
$\Omega_{\{r_{m-1}, r_m\}}
=\{z=r e^{\im \theta} \in \Omega : r_{m-1} \leq r < r_m\}$
with $r_m=e^{x_m}, 1 \leq m \leq M$,
and the boundary lines $\{z \in R: \Re z=x_m \}$
are to the arcs of semicircles
$\{z \in \Omega : |z|=r_m\}, 1 \leq m \leq M$.

\begin{figure}
\includegraphics[width=1.0\linewidth]{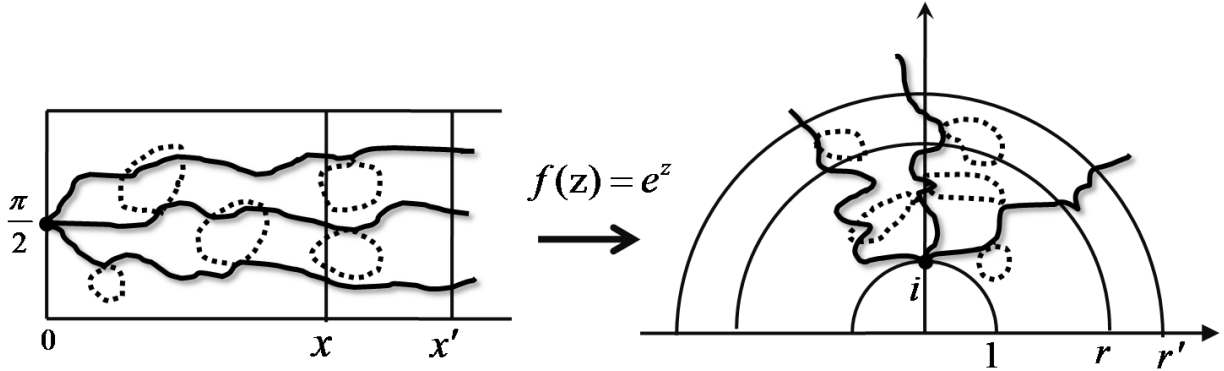}
\caption{Conformal transformation by $f(z)=e^z$
from the domain $R=\{z \in \C: \Re z >0,
0 < \Im z < \pi\}$
to the domain $\Omega=\{z \in \C: |z| > 1, 0< {\rm arg} (z) < \pi\}$.
The $N$-tuples of Brownian paths
in $R$
all starting from the point $\im \pi/2$
are conformally transformed into the paths
in $\Omega$ all starting from the point $\im$.
}
\label{fig:Fig6}
\end{figure}

By this conformal transformation,
the paths of complex Brownian motions in $R$ all starting from
the point $\im \pi/2$ are mapped to
those in $\Omega$ all starting from the point $\im$
as shown by Fig.\ref{fig:Fig6}.

The conformal invariance of the 
probability law of complex Brownian motions
(\ref{eqn:conformal}) implies the following equality
between the multiple correlation functions
$\rho_N^{\pi/2}$ defined on $R$ 
and $\hat{\rho}_N^{\im}$ defined on $\Omega$,
\begin{equation}
\rho_N^{\pi/2}(x_1, \vtheta^{(1)}_{N_1}; \dots; 
x_M, \vtheta^{(M)}_{N_M}) \prod_{m=1}^{M} d \vtheta^{(m)}_{N_m}
= \hat{\rho}_N^{\im}(\w^{(1)}_{N_1}; \dots; \w^{(M)}_{N_M})
\prod_{m=1}^{M} d \w^{(m)}_{N_m},
\label{eqn:conf2}
\end{equation}
where $\w^{(m)}_{N_m}=(w^{(m)}_1, \dots, w^{(m)}_{N_m})$
with $w^{(m)}_j= f(x_m+\im \theta^{(m)}_j)
=e^{x_m+\im \theta^{(m)}_j}=r_m e^{\im \theta^{(m)}_j}$,
$r_m \equiv e^{x_m}$, 
$1 \leq m \leq M$.
Since $0<x_1 < \cdots < x_M < \infty$ are fixed,
$$
dw^{(m)}_j=\left|
\frac{dw^{(m)}_j}{d \theta^{(m)}_j} \right|
d \theta^{(m)}_j
=r_m d \theta^{(m)}_j,
\quad 1 \leq m \leq M, 1 \leq j \leq N_m,
$$
we have the following
determinantal correlations for
first passage points on the semicircles in $\Omega$.

\vskip 0.5cm
\noindent{\bf Corollary 2.} \quad
Any multiple correlation function
in $\Omega$ is given by a determinant
\begin{equation}
\hat{\rho}_N^{\im}(\{r_1 e^{\im \theta^{(1)}_j} \}_{j=1}^{N_1};
\dots; \{r_M e^{\im \theta^{(M)}_j} \}_{j=1}^{N_M}) = 
\det_{1 \leq j \leq N_m, 1 \leq k \leq N_n, 
1 \leq m, n \leq M}
\Big[ \hat{\K}_N^{\im}(r_m e^{\im \theta^{(m)}_j},
r_n e^{\im \theta^{(n)}_k}) \Big]
\label{eqn:DeterminantalB1}
\end{equation}
with the correlation kernel
\begin{eqnarray}
&& \hat{\K}^{\im}_N(r e^{\im \theta}, r' e^{\im \theta'})
= \frac{2}{\pi r} \sum_{n=1}^N 
\frac{(r')^n-(r')^{-n}}{r^n-r^{-n}}
\sin(n \theta) \sin(n \theta'), \quad \mbox{if} \quad r \leq r', 
\nonumber\\
&& \hat{\K}_N^{\im}(r e^{\im \theta}, r' e^{\im \theta'})
= -\frac{2}{\pi r} \sum_{n=N+1}^{\infty}
\frac{(r')^n-(r')^{-n}}{r^n-r^{-n}}
\sin(n\theta) \sin(n\theta'), \quad \mbox{if} \quad r >r',
\label{eqn:hatK}
\end{eqnarray}
$1 < r, r' < \infty, 0 < \theta, \theta' < \pi$.
\vskip 0.5cm

\begin{figure}
\includegraphics[width=0.8\linewidth]{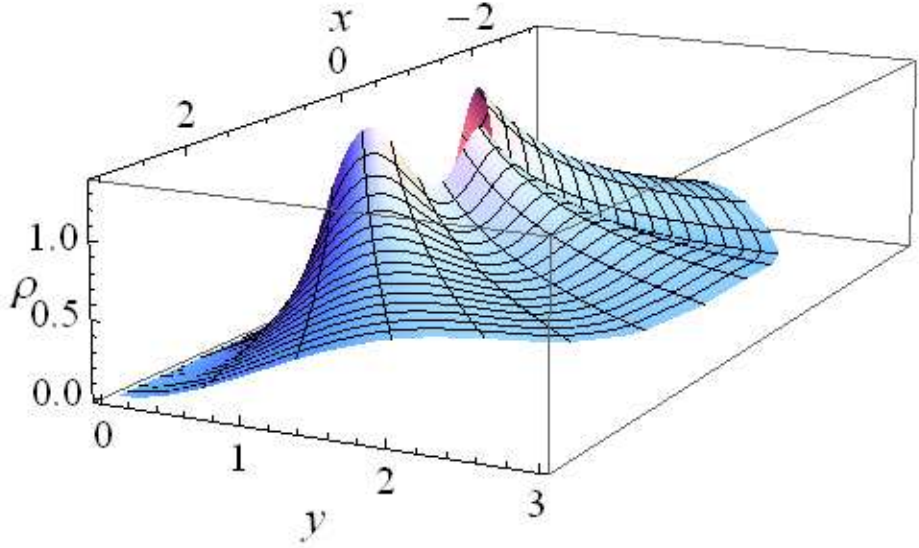}
\caption{(Color online)
The density function
$\hat{\rho}_3^{\im}(r e^{\im \theta})$ for $N=3$,
where $x=\Re(r e^{\im \theta})=r \cos \theta$, 
$y=\Im(r e^{\im \theta})=r \sin \theta$.
There are three ridges.
}
\label{fig:Fig7}
\end{figure}

From (\ref{eqn:hatK}), we find that if we set $r=r'>1$,
\begin{eqnarray}
&& \hat{\K}^{\im}_N(re^{\im \theta}, r e^{\im \theta'})
= \frac{2}{\pi r} \sum_{n=1}^{N}
\sin(n \theta) \sin(n \theta')
\nonumber\\
&& \quad = 
\frac{\sin((N+1)\theta) \sin N\theta' - \sin N\theta \sin((N+1)\theta')}
{\pi r (\cos\theta - \cos\theta')}
\label{eqn:KN0}
\end{eqnarray}
for $\theta \not= \theta'$. From it 
the density function 
$\hat{\rho}_N^{\im}(r e^{\im \theta})=\lim_{\varepsilon \to 0}
\hat{\K}^{\im}_N(re^{\im \theta}, r e^{\im (\theta+\varepsilon)})
$
is given as
\begin{equation}
\hat{\rho}_N^{\im}(r e^{\im \theta})= \frac{1}{\pi r \sin \theta}
\Big[ N \sin \theta-\cos \theta \cos(N \theta) \sin (N \theta)
+\sin \theta \sin^2(N \theta) \Big].
\label{eqn:rho0}
\end{equation}
For $N=3$, 
Fig.\ref{fig:Fig7} shows the dependence
of $\hat{\rho}_3^{\im}(r e^{\im \theta})$
on $x=\Re(r e^{\im \theta})=r \cos \theta$
and $y=\Im(r e^{\im \theta})=r \sin \theta$.
There are $N=3$ ridges in the plots.

\begin{figure}
\includegraphics[width=0.6\linewidth]{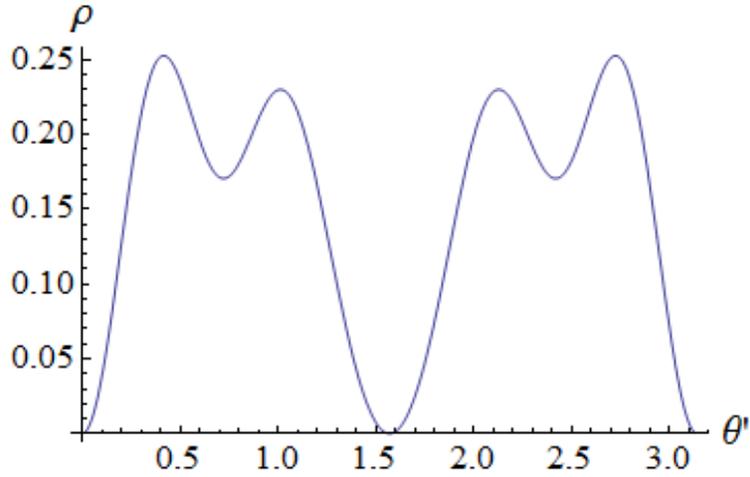}
\caption{For $N=5$, the two-point correlation function 
(\ref{eqn:two1}) on an arc
of semicircle $|z|=r=4$ with $\theta=\pi/2$
is shown as a function of $\theta'$.
There are $N-1=4$ peaks in the plot.
}
\label{fig:Fig8}
\end{figure}

\begin{figure}
\includegraphics[width=0.6\linewidth]{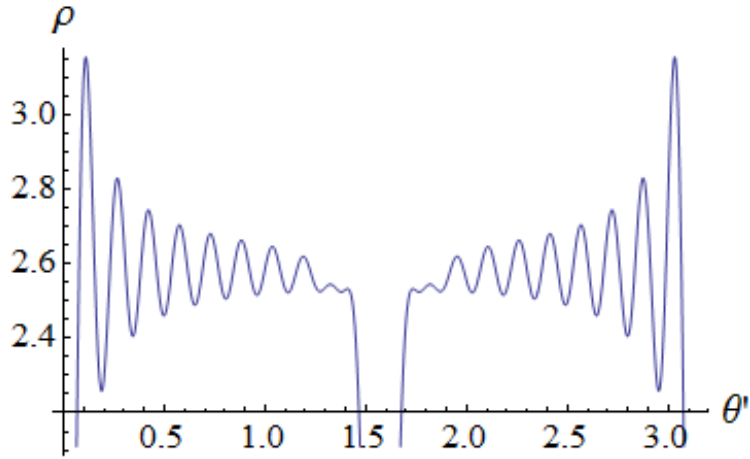}
\caption{For $N=20$, the two-point correlation function 
(\ref{eqn:two1}) on an arc
of semicircle $|z|=r=4$ with $\theta=\pi/2$
is shown as a function of $\theta'$.
}
\label{fig:Fig9}
\end{figure}

On an arc of semicircle $|z|=r >1, 0 < {\rm arg} (z) < \pi$,
the two-point correlation function is given by
\begin{equation}
\hat{\rho}^{\im}_N(r e^{\im \theta}, r e^{\im \theta'})
=\hat{\rho}^{\im}_N(r e^{\im \theta})
\hat{\rho}^{\im}_N(r e^{\im \theta'})
-(\hat{\K}^{\im}_N(r e^{\im \theta}, r e^{\im \theta'}))^2
\label{eqn:two1}
\end{equation}
with (\ref{eqn:KN0}) and (\ref{eqn:rho0})
for $0 < \theta, \theta' < \pi$.
In Figs.\ref{fig:Fig8} and \ref{fig:Fig9}, 
we set $r=4$ and $\theta=\pi/2$ and plot
(\ref{eqn:two1}) as a function of $\theta'$
for $N=5$ and $N=20$, respectively.
Due to the nonintersection condition
for loop-erased parts,
the two-point correlation function
becomes zero as $\theta' \to \theta=\pi/2$.

\begin{figure}
\includegraphics[width=0.6\linewidth]{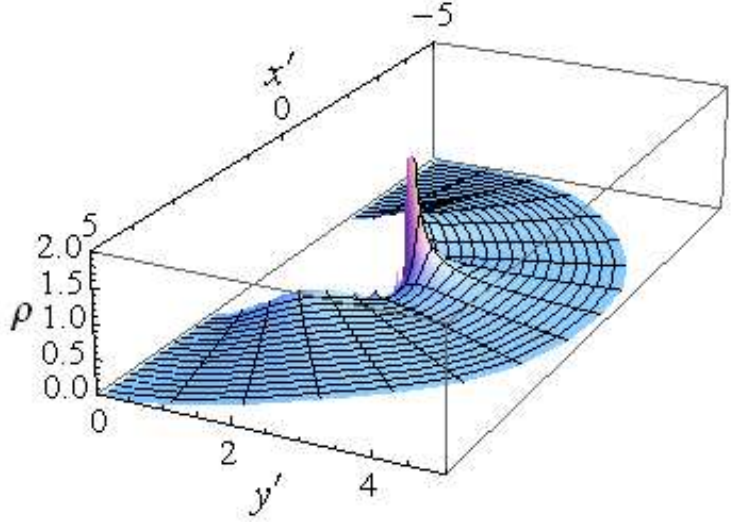}
\caption{(Color online)
Two-point correlation function
(\ref{eqn:two2}) with $N=3, r=2, \theta=\pi/2$
is shown as a function of
$x'=\Re(r' e^{\im \theta'})=r' \cos \theta'$
and 
$y'=\Re(r' e^{\im \theta'})=r' \sin \theta'$.
}
\label{fig:Fig10}
\end{figure}

In general, 
for $1 < r < r', 0 < \theta, \theta' < \pi$,
Corollary 2 gives the two-point correlation function as
\begin{eqnarray}
&& \hat{\rho}^{\im}_N(r e^{\im \theta}, r' e^{\im \theta'})
=\hat{\rho}^{\im}_N(r e^{\im \theta})
\hat{\rho}^{\im}_N(r' e^{\im \theta'})
+ \frac{4}{\pi^2 r r'}
\sum_{n=1}^{N} 
\frac{(r')^n-(r')^{-n}}{r^n-r^{-n}}
\sin(n \theta) \sin(n \theta')
\nonumber\\
&& \qquad \qquad \times
\sum_{m=N+1}^{\infty}
\frac{r^m-r^{-m}}{(r')^m-(r')^{-m}}
\sin(m \theta') \sin(m \theta)
\label{eqn:two2}
\end{eqnarray}
with (\ref{eqn:rho0}).
For $N=3$ we set
$r e^{\im \theta}=2 e^{\im \pi/2}=2 \im$
and show by Fig.\ref{fig:Fig10} 
the dependence of the two-point correlation
function (\ref{eqn:two2}) on 
$x'=\Re(r' e^{\im \theta'})=r' \cos \theta'$
and 
$y'=\Re(r' e^{\im \theta'})=r' \sin \theta'$.

\section{Concluding Remarks}

Here we discuss 
the infinite number of paths
limit, $N \to \infty$.
When we take this limit in (\ref{eqn:rho0}), we have
\begin{equation}
\lim_{N \to \infty} \frac{1}{N} \hat{\rho}_N^{\im}
(r e^{\im \theta}) 
=\frac{1}{\pi r}.
\label{eqn:uniform}
\end{equation}
That is, the distribution of the first passage point 
becomes uniform on an arc of semicircle.
If we set
$$
r=N+u, \quad r'=N+u', \quad
\theta=\frac{a}{N}, \quad \theta'=\frac{a'}{N},
$$
then the correlation kernel converges to the 
following as $N \to \infty$,
\begin{eqnarray}
&& \hat{\K}^{\im}(u,a; u', a')
= \lim_{N \to \infty} 
\hat{\K}^{\im}_N((N+u) e^{\im a/N}, (N+u') e^{\im a'/N}))
\nonumber\\
&& \quad = \left\{ \begin{array}{ll}
\displaystyle{
\frac{2}{\pi} \int_0^1 e^{-(u-u')s} \sin(as) \sin(as') ds},
& \quad \mbox{if} \quad u<u', \\
\displaystyle{
-\frac{2}{\pi} \int_1^{\infty} e^{-(u-u')s} \sin(as) \sin(as') ds},
& \quad \mbox{if} \quad u > u'.
\end{array} \right.
\label{eqn:limitK}
\end{eqnarray}

In the present paper, we have imposed special
initial conditions such that all Brownian paths
start from a single point $\im \pi/2$ for
the domain $R$ and from $\im$ for the domain $\Omega$.
Study for general initial condition will be reported
elsewhere in the future.

At the end of this paper, we note the fact that
the scaling limit of LERW is described by the SLE(2) path,
a random continuous simple curve generated by the
Schramm-Loewner evolution 
with a special value of parameter $\kappa=2$ \cite{Sch00,LSW04,Law05}.
Kozdron \cite{Koz09} showed that $2 \times 2$ Fomin's determinant
representing the event $\LE(\gamma_1) \cap \gamma_2=\emptyset$
for two Brownian paths $(\gamma_1, \gamma_2)$
is proportional to the probability that 
$\gamma_{\rm SLE(2)} \cap \gamma=\emptyset$,
where $\gamma_{\rm SLE(2)}$ and $\gamma$ denote the SLE(2) path
and a Brownian path (see also \cite{KL07}).
On the other hand, Lawler and Werner gave a method
to correctly add Brownian loops to an SLE(2) path
to obtain a Brownian path \cite{LW04}.
Interpretation of the results reported in the present paper
in terms of `mutually avoiding SLE paths' will be
an interesting future problem.

\begin{acknowledgments}
The present authors would like to thank 
Michael Kozdron for careful reading of the manuscript
and for useful comments on mathematics
of loop-erased random walks, their continuum limits
and SLE.
M.K. is supported in part by
the Grant-in-Aid for Scientific Research (C)
(No.21540397) of Japan Society for
the Promotion of Science.
\end{acknowledgments}

\appendix
\section{Derivation of $H_{R_L}$ and $H_{\partial R_L}$}

For $z=x+\im y \in R_L, L >0$, we solve the Laplace equation
\begin{equation}
\left( \frac{\partial^2}{\partial x^2}
+\frac{\partial^2}{\partial y^2} \right) 
H_{R_L}(x+\im y, L+ \im \rho)=0
\label{eqn:A1}
\end{equation}
for $0 < \rho < \pi$, by the method of
separation of variables.
We set 
$H_{R_L}(x+\im y, L+\im \rho)=X(x) Y(y)$,
where description of dependence
on $L$ and $\rho$ is omitted.
Then we have a pair of ordinary differential equations
\begin{eqnarray}
\label{eqn:A2}
&& X''(x)=c X(x), \\
\label{eqn:A3}
&& Y''(y)=-c Y(y)
\end{eqnarray}
with a constant $c$, which does not depend on $x$ nor $y$.
With the boundary condition
$Y(0)=Y(\pi)=0$, Eq.(\ref{eqn:A3}) is solved as
\begin{equation}
Y(y)=a \sin(ny), \quad
\sqrt{c}=n, \quad n \in \N,
\quad 0 < y < \pi
\label{eqn:A4}
\end{equation}
with a constant $a$.
Then Eq.(\ref{eqn:A2}) becomes
$X''(x)=n^2 X(x)$, which is solved under the
condition $X(0)=0$ as
\begin{equation}
X(x)=b \sinh (nx), \quad
0 < x < L.
\label{eqn:A5}
\end{equation}
Then we have the form
\begin{equation}
H_{R_L}(x+\im y, L+\im \rho)
=\sum_{n=1}^{\infty} c_n(L, \rho)
\sinh(nx) \sin(ny)
\label{eqn:A6}
\end{equation}
with a series of coefficients $\{c_n(L, \rho) \}$, where
dependence on $L$ and $\rho$ is now revealed.
In this case
the boundary condition for the Poisson kernel
(\ref{eqn:Poisson2}) becomes
\begin{equation}
\lim_{x \to L} H_{R_L}(x+\im y, L+\im \rho)
=\delta(y-\rho),
\label{eqn:A7}
\end{equation}
which uniquely determines the coefficients as
\begin{equation}
c_n(L, \rho)=\frac{2}{\pi}
\frac{\sin(n \rho)}{\sinh(nL)},
\label{eqn:A8}
\end{equation}
since the Fourier series of 
the Dirac delta function is known as
$$
\delta(y-\rho)=\frac{2}{\pi}
\sum_{n=1}^{\infty} \sin(ny) \sin(n \rho)
$$
for $y, \rho >0$.

Following (\ref{eqn:Poisson3}),
the boundary Poisson kernel is obtain by
taking the limit as
\begin{eqnarray}
&& H_{\partial R_L}(\im y, L+\im \rho)
= \lim_{\varepsilon \to 0}
\frac{1}{\varepsilon}
H_{R_L}(\im y+ \varepsilon, L+\im \rho)
\nonumber\\
&& \quad =
\frac{2}{\pi} \sum_{n=1}^{\infty}
\frac{n \sin(ny) \sin(n \rho)}{\sinh(nL)}.
\nonumber
\end{eqnarray}



\end{document}